\documentclass[12pt]{article}
 \usepackage{epsfig}
 \def\be{\begin{equation}}
 \def\ee{\end{equation}}
 \def\bea{\begin{eqnarray}}
 \def\eea{\end{eqnarray}}
 \usepackage{graphicx}

 \catcode`\@=11
 \def\lsim{\mathrel{\mathpalette\@versim<}}
 \def\gsim{\mathrel{\mathpalette\@versim>}}
 \def\@versim#1#2{\vcenter{\offinterlineskip
 \ialign{$\m@th#1\hfil##\hfil$\crcr#2\crcr\sim\crcr } }}
 \catcode`\@=12

 \catcode`@=12
\begin{document}
 \thispagestyle{empty}
 \begin{flushright}
 UCRHEP-T616\\
 May 2021\
 \end{flushright}
 \vspace{0.6in}
 \begin{center}
 {\LARGE \bf New $[SU(3)]^4$ Realization of\\
Lepton/Dark Symmetry\\}
 \vspace{1.5in}
 {\bf Ernest Ma\\}
 \vspace{0.1in}
{\sl Department of Physics and Astronomy,\\ 
University of California, Riverside, California 92521, USA\\}
\end{center}
 \vspace{1.2in}

\begin{abstract}\
Extending the well-known $SU(3)_C \times SU(3)_L \times SU(3)_R$ model of 
quarks and leptons to include a fourth $SU(3)_N$ gauge factor, a new 
realization is obtained, different from leptonic color, which contains 
a lepton/dark symmetry with the help of an input $Z_4$ symmetry.  It is 
seen to encompass a previous extension of the standard model to $SU(2)_N$ 
lepton symmetry.
\end{abstract}

\newpage
\baselineskip 24pt
\noindent \underline{\it Introduction}~:~ 
It is well-known that the standard model (SM) of quarks and leptons may be 
embedded in $SU(5)$ with the fermions as $\underline{5}^*$ and 
$\underline{10}$ representations per family.  Adding the right-handed 
neutrino, they form a single $\underline{16}$ representation of $SO(10)$. 
It is also well-known that this $\underline{16}$ may be embedded in the 
$\underline{27}$ of $E_6$.  This last set of fermions has an interesting 
realization, using the maximal $SU(3)_C \times SU(3)_L \times SU(3)_R$ 
subgroup of $E_6$, i.e.
\begin{equation}
q \sim (3,3^*,1) \sim \pmatrix{d & u & h \cr d & u & h \cr d & u & h}, ~~~  
 q^c \sim (3^*,1,3) \sim \pmatrix{d^c & d^c & d^c \cr 
u^c & u^c & u^c \cr h^c & h^c & h^c},
\end{equation}
\begin{equation}
f \sim (1,3,3^*) \sim \pmatrix{N_1^0 & E^c & \nu \cr E & N_2^0 & e \cr 
\nu^c & e^c & S^0},
\end{equation}
where
\begin{equation}
Q = I_{3L} - {1 \over 2} Y_L + I_{3R} - {1 \over 2} Y_R.
\end{equation}
The columns in the $3 \times 3$ fermion matrices denote $3$ representations 
of $SU(3)$ with $(I_3,Y) = (1/2,1/3),(-1/2,1/3),(0,-2/3)$ from top to bottom. 
The rows denote $3^*$ representations with 
$(I_3,Y) = (-1/2,-1/3),(1/2,-1/3),(0,2/3)$ from left to right.  The scalar 
\begin{equation}
\lambda_0 \sim (1,3,3^*) \sim \pmatrix{\eta_1^0 & \eta_2^+ & \phi_L^0 
\cr \eta_1^- & \eta_2^0 & \phi_L^- \cr \phi_R^0 & \phi_R^+ & \zeta^0}
\end{equation}
transforms as $f$, with allowed Yukawa couplings
\begin{equation}
Tr[q \lambda_0 q^c], ~~~ \epsilon_{abc} \epsilon_{\alpha \beta \gamma} 
f_{a \alpha} f_{b \beta} (\lambda_0)_{c \gamma}.
\end{equation}

To generalize the above, $[SU(N)]^k$ in a moose chain, i.e. fermions of the 
form $(N,N^*,1...1)$, $(1,N,N^*,...1),...$ to $k$ copies, may be considered. 
In general, supersymmetric $[SU(N)]^k$ has the intriguing property that 
it is a finite field theory~\cite{mmz04} with three families, for any $N$ 
or $k$.  For $N=3$ and $k=4$, the fourth 
$SU(3)$~\cite{bmw04,kmppz17,kmppz18,kmppz18-1} could be leptonic 
color~\cite{fl90,flv91} or not~\cite{m05}.  For $N=3$ and $k=6$, 
the fermions of $[SU(3)]^3$ are separated by three additional $SU(3)$ 
factors~\cite{m04,ccmr06} to allow for chiral color~\cite{z83,fg87}.

In this paper, a new choice of the fourth $SU(3)$ is studied for an 
$[SU(3)]^4$ model.  It will be shown that it contains a previously 
proposed~\cite{fstw17,m21} $SU(2)_N$ lepton symmetry.  Furthermore, 
residual conserved global baryon number $B$ and lepton number $L$ 
may be defined, and dark symmetry is derivable~\cite{m15,m20} from 
lepton symmetry, with vector gauge bosons~\cite{dm11,bdmw12,mw12} 
in the dark sector~\cite{bh18}.

\noindent \underline{\it Model}~:~ 
The gauge symmetry is $SU(3)_C \times SU(3)_L \times SU(3)_N \times SU(3)_R$. 
The fermions transform as
\begin{eqnarray}
&& q \sim (3,3^*,1,1) \sim \pmatrix{d & u & h \cr d & u & h \cr d & u & h},  
~~~ l \sim (1,3,3^*,1) \sim \pmatrix{N_1 & \nu & E_1^c \cr E_1 & e & N_1^c 
\cr S_2 & S_1 & E_0^c}, \\ 
&& l^c \sim (1,1,3,3^*) \sim \pmatrix{N_2^c & E_2^c & S_2^c \cr 
\nu^c & e^c & S_1^c \cr E_2 & N_2 & E_0},  
~~~ q^c \sim (3^*,1,1,3) \sim \pmatrix{d^c & d^c & d^c \cr u^c & u^c & u^c 
\cr h^c & h^c & h^c},
\end{eqnarray}
with the electric charge given by
\begin{equation}
Q = I_{3L} - {1 \over 2} Y_L + I_{3R} - {1 \over 2} Y_R + Y_N.
\end{equation}
Again, the columns in the $3 \times 3$ fermion matrices denote $3$ 
representations of $SU(3)$ with $(I_3,Y) = (1/2,1/3),(-1/2,1/3),(0,-2/3)$ 
from top to bottom.  The rows denote $3^*$ representations with 
$(I_3,Y) = (-1/2,-1/3),(1/2,-1/3),(0,2/3)$ from left to right. 
The $SU(2)_N$ of Refs.\cite{fstw17,m21} is clearly embedded in $l,l^c$.

The scalars transform as
\begin{equation}
\lambda_0 \sim (1,3,1,3^*) \sim \pmatrix{\eta_1^0 & \eta_2^+ & \phi_L^0 
\cr \eta_1^- & \eta_2^0 & \phi_L^- \cr \phi_R^0 & \phi_R^+ & \zeta^0}, 
\end{equation}
\begin{equation}
\lambda_L \sim (1,3,3^*,1) \sim \pmatrix{\phi_2^0 & \phi_1^0 & \phi_3^+ 
\cr \phi_2^- & \phi_1^- & \phi_3^0 \cr \chi_2 & \chi_1 & \zeta_L^+},  
~~~ \lambda_R \sim (1,1,3,3^*) \sim \pmatrix{\phi_4^0 & \phi_4^+ & \chi_4 
\cr \phi_5^0 & \phi_5^+ & \chi_5 \cr \phi_6^- & \phi_6^0 & \zeta_R^-}.
\end{equation}
The allowed Yukawa terms are
\begin{equation}
Tr[q^c q \lambda_0], ~~~ Tr[\lambda_0^\dagger l l^c], ~~~ \epsilon_{abc} 
\epsilon_{\alpha \beta \gamma} l_{a \alpha} l_{b \beta} 
(\lambda_L)_{c \gamma}, ~~~ \epsilon_{abc} \epsilon_{\alpha \beta \gamma} 
l^c_{a \alpha} l^c_{b \beta} (\lambda_R)_{c \gamma}.
\end{equation}
Two other scalars are added: $\lambda'_L \sim (1,3,3^*,1)$ and 
$\lambda'_R \sim (1,1,3,3^*)$, together with a $Z_4$ symmetry. 
Under $Z_4$,
\begin{equation}
q,q^c,\lambda_0 \sim 1, ~~~ \lambda_{L,R} \sim -1, ~~~ l,\lambda'_L \sim i, 
~~~ l^c,\lambda'_R \sim -i.
\end{equation}
Hence Eq.~(11) remains valid and $\lambda'_{L,R}$ do not couple to $l,l^c$. 
Allowed trilinear scalar couplings are 
$Tr[\lambda_0^\dagger \lambda_L \lambda_R]$ and 
$Tr[\lambda_0^\dagger \lambda'_L \lambda'_R]$, 
$\epsilon_{abc} \epsilon_{\alpha \beta \gamma} (\lambda_0)_{a \alpha} 
(\lambda_0)_{b \beta} (\lambda_0)_{c \gamma}$, 
$\epsilon_{abc} \epsilon_{\alpha \beta \gamma} (\lambda'_L)_{a \alpha} 
(\lambda'_L)_{b \beta} (\lambda_L)_{c \gamma}$, and 
$\epsilon_{abc} \epsilon_{\alpha \beta \gamma} (\lambda'_R)_{a \alpha} 
(\lambda'_R)_{b \beta} (\lambda_R)_{c \gamma}$.  Note that cubic terms 
of the form $(\lambda_L)^3$, $(\lambda_R)^3$, $(\lambda'_L)^3$, or 
$(\lambda'_R)^3$ are all forbidden by $Z_4$.  The absence of these terms 
will lead to a residual lepton/dark symmetry as discussed in the next section.

\noindent \underline{\it Residual $B$ and $L$ Symmetries}~:~ 
Under $SU(3)_L \times SU(3)_N \times SU(3)_R$, the following neutral scalars 
have vacuum expectation values (VEVs), as shown in Table~1.
\begin{table}[tbh]
\centering
\begin{tabular}{|c|c|c|c|c|c|c|c|}
\hline
scalar & $I_{3L}$ & $Y_L$ & $I_{3N}$ & $Y_N$ & $I_{3R}$ & $Y_R$ & VEV \\
\hline
$\zeta^0$ & $0$ & $-2/3$ & $0$ & $0$ & $0$ & $2/3$ & $u_0$ \\ 
$\eta_1^0$ & $1/2$ & $1/3$ & $0$ & $0$ & $-1/2$ & $-1/3$ & $v_1$ \\ 
$\eta_2$ & $-1/2$ & $1/3$ & $0$ & $0$ & $1/2$ & $-1/3$ & $v_2$ \\ 
$\phi_L^0$ & $1/2$ & $1/3$ & $0$ & $0$ & $0$ & $2/3$ & $v_L$ \\ 
$\phi_R^0$ & $0$ & $-2/3$ & $0$ & $0$ & $-1/2$ & $-1/3$ & $v_R$ \\ 
\hline
$\chi_1$ & $0$ & $-2/3$ & $1/2$ & $-1/3$ & $0$ & $0$ & $u_L$ \\
$\chi_5$ & $0$ & $0$ & $-1/2$ & $1/3$ & $0$ & $2/3$ & $u_R$ \\
\hline
${\phi_3^0}'$ & $-1/2$ & $1/3$ & $0$ & $2/3$ & $0$ & $0$ & $v_3$ \\
${\phi_6^0}'$ & $0$ & $0$ & $0$ & $-2/3$ & $1/2$ & $-1/3$ & $v_6$ \\
\hline
\end{tabular}
\caption{Scalars with vacuum expectation values.}
\end{table}

From the allowed Yukawa couplings
\begin{equation}
d^d d \eta_1^0, ~~~ u^c u \eta_2^0, ~~~ h^c h \zeta^0, ~~~ d^c h \phi_L^0, 
~~~ h^c d \phi_R^0,
\end{equation}
it is seen that the $u$ quarks get masses from $v_2$, and the $d,h$ quarks 
get diagonal masses from $v_1$ and $u_0$, with mixing terms from $v_{L,R}$. 
The $l$ and $l^c$ fermions have the allowed Yukawa couplings
\begin{eqnarray}
&& (\nu \nu^c + N_1 N_2^c + E_1^c E_2) \bar{\eta}_1^0, ~~~ 
(e e^c + N_1^c N_2 + E_1 E_2^c) \bar{\eta}_2^0, ~~~ 
(S_1 S_1^c + S_2 S_2^c + E_0^c E_0) \bar{\zeta}^0, \\ 
&& (\nu S_1^c + N_1 S_2^c + E_1^c E_0) \bar{\phi}_L^0, ~~~ 
(S_1 \nu^c + S_2 N_2^c + E_0^c E_2) \bar{\phi}_R^0, \\
&& (E_1 E_1^c - N_1 N_1^c) \chi_1, ~~~ (E_2 E_2^c - N_2 N_2^c) \chi_5.
\end{eqnarray}
It is seen that the charged leptons get masses from $v_2$, whereas $S_{1,2}$ 
and $E_0$ get masses from $u_0$, $(N_1,E_1)$ from $u_L$, $(N_2,E_2)$ from 
$u_R$, with mixing among them (except $S_1$) from $v_{L,R}$ as well as 
$v_{1,2}$.  The neutrinos have diagonal masses from $v_1$, but they also 
mix with $S_1$ from $v_{L,R}$.  Note that $v_{3,6}$ are not involved in 
fermion masses because $\lambda'_{L,R}$ do not couple to $l,l^c$.

From the above, two global residual symmetries may be defined. (1) Baryon 
number $B=1/3$ for $q$ and $B=-1/3$ for $q^c$. (2) Lepton number $L=1$ for 
$\nu,e,S_1$, and $L=-1$ for $\nu^c,e^c,S_1^c$.  Under $L$, 
\begin{eqnarray}
&& l \sim \pmatrix{0 & 1 & 0 \cr 0 & 1 & 0 \cr 0 & 1 & 0}, ~~~ 
l^c \sim \pmatrix{0 & 0 & 0 \cr -1 & -1 & -1 \cr 0 & 0 & 0}, \\  
&& \lambda_L \sim \pmatrix{-1 & 0 & -1 \cr -1 & 0 & -1 \cr -1 & 0 & -1}, ~~~ 
\lambda_R \sim \pmatrix{-1 & -1 & -1 \cr 0 & 0 & 0 \cr -1 & -1 & -1}.
\end{eqnarray}
The $\lambda_0$ scalars all have $L=0$. 

It is also clear that the complex vector gauge bosons in $SU(3)_N$ 
which take the $(1/2,-1/3)$ state under $(I_{3N},Y_N)$ to the $(-1/2,-1/3)$ 
and $(0,2/3)$ states must have $L=-1$.  For the $\lambda'_{L,R}$ scalars, 
${\phi_3^0}'$ and ${\phi_6^0}'$ are required to have $L=0$ for them to 
acquire VEVs.  Together with the nonzero $L$ assignments in the gauge 
sector, $\lambda'_{L,R} \sim l,l^c$ under $L$ is then fixed.  The input 
$Z_4$ symmetry is thus responsible for $L$ conservation in this model 
including $\lambda'_{L,R}$.

If $\lambda'_{L,R}$ are absent, then the extra $Z_4$ symmetry is not 
necessary.  Fermion masses are unaffected because they depend only on 
$u_{0,L,R}$ and $v_{1,2,L,R}$, but without $v_{3,6}$, the symmetry breaking 
of $SU(3)_L \times SU(3)_N \times SU(3)_R$ would not be realistic.  
Specifically, $v_3$ breaks $SU(3)_N$ and $SU(2)_L$, and $v_6$ breaks 
$SU(3)_N$ and $SU(2)_R$.

\noindent \underline{\it Gauge Sector}~:~
The $SU(3)_C$ gauge factor contains the gluon octet.  Each of the other 
$SU(3)$ factors contains eight vector gauge bosons, i.e.
\begin{equation}
{g \over 2} \pmatrix{W_3 + W_8/\sqrt{3} & W_1-iW_2 & W_4-iW_5 \cr 
W-1+iW_2 & -W_3 + W_8/\sqrt{3} & W_6-iW_7 \cr W_4+iW_5 & W_6+iW_7 & 
-2W_8/\sqrt{3}}.
\end{equation}
Of the nine VEVs, four ($v_{1,2,3,L}$) contribute to the mass of $W_{3L}$. 
They must be small compared to the other five VEVs, from which four 
of the five vector fields $(W_{3N},W_{3R},W_{8L},W_{8N},W_{8R})$ obtain mass. 
If $v_6$ is missing, only three would do so.  As it is, one linear 
combination is the analog of the $U(1)_Y$ gauge boson of the SM and 
would get a mass from $v_{1,2,3,L}$.  It mixes with $W_{3L}$ to form 
the photon and the SM $Z$ boson in the usual way.  In the limit 
$g_L=g_R=g_N$, this state is given by
$(\sqrt{3} W_{3R} + 2 W_{8N} - W_{8L} - W_{8R})/3$.
 
Under $SU(3)_N$, the gauge boson $(W_{1N}-iW_{2N})/\sqrt{2}$ has $L=1$ and 
$Q=0$, $(W_{4N}-iW_{5N})/\sqrt{2}$ has $L=0$ and $Q=1$, and 
$(W_{6N}-iW_{7N})/\sqrt{2}$ has $L=-1$ and $Q=1$.  Their respective 
masses are $g_N \sqrt{(u_L^2+u_R^2)/2}$, $g_N \sqrt{(v_3^2+v_6^2)/2}$, and 
$g_N \sqrt{(u_L^2+u_R^2+v_3^2+v_6^2)/2}$.  To summarize,
\begin{equation}
L(W_N) \sim \pmatrix{0 & 1 & 0 \cr -1 & 0 & -1 \cr 0 & 1 & 0}, ~~~ 
Q(W_N) \sim \pmatrix{0 & 0 & 1 \cr 0 & 0 & 1 \cr -1 & -1 & 0}, 
\end{equation}
whereas 
\begin{equation}
L(W_{L,R}) \sim 0, ~~~ 
Q(W_{L,R}) \sim \pmatrix{0 & 1 & 0 \cr -1 & 0 & -1 \cr 0 & 1 & 0}. 
\end{equation}

\noindent \underline{\it Dark Sector}~:~
With the conservation of lepton number $L$ as defined in the previous 
sections, a dark parity may be derived~\cite{m15}, i.e. $\pi_D = (-1)^{L+2j}$. 
This means that $\nu,e,S_1$ are even, but $(N,E)_{1,2},S_2,E_0$ are odd. 
The scalars in $\lambda_0$ are even, together with 
$\Phi_1, \chi_1, \Phi_5, \chi_5$, in $\lambda_{L,R}$, whereas 
$\Phi_2, \chi_2, \Phi_3, \Phi_4, \chi_4, \Phi_6, \zeta^\pm$ are odd.  
The scalars in $\lambda'_{L,R}$ have the opposite $\pi_D$ as the ones 
in $\lambda_{L,R}$.  The vector gauge bosons are all even except for 
$(W_{1N} \pm iW_{2N})/\sqrt{2}$ and $(W_{6N} \pm iW_{7N})/\sqrt{2}$ which 
are odd.  The neutral $(W_{1N} \pm iW_{2N})/\sqrt{2}$ may be 
dark matter and is analogous to the $X$ boson in Ref.~\cite{m21}.  
Similarly, a linear combination of $\chi_{1,5}$ is analogous to the $H$ 
scalar boson discussed there.  The $(N_1,E_1)$ and $(N_2,E_2)$ fermions 
are analogous to $(N,E)$ and $(N',E')$ respectively.  The dark-matter 
phenomenology is thus the same.

\noindent \underline{\it Phenomenology of $SU(3)_N$}~:~
The breaking of $SU(3)_{L,R}$ to $SU(2)_{L,R}$ is assumed to be at a 
high scale.  The subsequent breaking of $SU(2)_R$ is also assumed to be 
high.  These may be accomplished by $u_0$ and $v_R$ as shown in Table 1. 
The $SU(3)_N$ breaking to $SU(2)_N$ is through $v_6$ which also breaks 
$SU(2)_R$.  Finally, $SU(2)_N$ breaking is through $u_{L,R}$.  This 
chain allows the neutral vector gauge boson $X=(W_{1N}-iW_{2N})/\sqrt{2}$ 
to be dark matter as in Ref.~\cite{m21}.

The charged leptons have interactions with the $W_{L,R,N}$ gauge bosons 
as well as the new fermions and scalars contained in $l,l^c$ and 
$\lambda_{0,L,R}$.  This means that there are many possible one-loop 
contributions to the muon anomalous magnetic moment, for example. 
For a study restricted to the simplified $SU(2)_N$ sector, see 
Ref.~\cite{m21}.

Since the $SU(3)_N$ gauge bosons couple only to $l,l^c$ fermions and 
$\lambda_{L,R}$, $\lambda'_{L,R}$ scalars, they are not easily produced. 
The highest energy of the $e^+e^-$ LEP II collider was 209 GeV.  Hence  
the $W_{3N}$ and $W_{8N}$ bosons should be heavier than this value. 

There are three $SU(2)_L$ scalar doublets $\Phi_{1,2,3}$ in  $\lambda_L$ 
and three $SU(2)_R$ scalar doublets $\Phi_{4,5,6}$ in  $\lambda_R$.  
They have different $L$ values as shown in Eq.~(17), and are connected 
by $(W_{1N}\pm iW_{2N})/\sqrt{2}$, $(W_{4N}\pm iW_{5N})/\sqrt{2}$, and 
$(W_{6N}\pm iW_{7N})/\sqrt{2}$.  A similar pattern exists also for 
$\lambda'_{L,R}$.  Hence this model predicts many more scalars beyond 
the lone Higgs boson of the SM.  

\noindent \underline{\it Concluding Remarks}~:~
A new realization of $[SU(3)]^4$ gauge symmetry is proposed, embedding 
the SM quarks and leptons as shown in Eqs.~(6) and (7).  The new $SU(3)_N$ 
symmetry has a neutral $SU(2)_N$ subgroup which identifies with the
non-Abelian lepton symmetry proposed before~\cite{fstw17,m21}.  It is 
shown how all fermions in $q,q^c,l,l^c$ may acquire mass with the breaking 
of $[SU(3)]^4$ to the SM gauge symmetry, then to $SU(3)_C \times U(1)_Q$. 
With the help of a $Z_4$ symmetry which applies to $q,q^c,l,l^c$ fermions 
and the $\lambda_{0,L,R}, \lambda'_{L,R}$ scalars, it is shown that two 
conserved residual symmetries remain.  One is the usual baryon number $B$; 
the other is generalized lepton number $L$, as shown in Eqs.~(16), (17), (19), 
and (20).  Hence two complex vector gauge bosons (one neutral and one 
charged) have $L \neq 0$.  The former may be dark matter, as discussed in 
Ref.~\cite{m21}, with dark parity $\pi_D = (-1)^{L+2j}$.  As most presumed 
candidates of dark matter are either scalar or fermion, this possibility 
should not be overlooked.

\noindent \underline{\it Acknowledgement}~:~
This work was supported 
in part by the U.~S.~Department of Energy Grant No. DE-SC0008541.

\bibliographystyle{unsrt}

\end{document}